\documentclass{emulateapj}
\usepackage{apjfonts}
\usepackage{graphicx}

\shortauthors{Sanchis-Ojeda et al.~2011}
\shorttitle{Spin-orbit misalignment of HAT-P-11}

\begin{document}

%
\def\ltsima{$\; \buildrel < \over \sim \;$}
\def\lsim{\lower.5ex\hbox{\ltsima}}
\def\gtsima{$\; \buildrel > \over \sim \;$}
\def\gsim{\lower.5ex\hbox{\gtsima}}
%

\bibliographystyle{apj}

\title{
  Starspots, spin-orbit misalignment, and active latitudes \\
   in the HAT-P-11 exoplanetary system
}

\author{
Roberto Sanchis-Ojeda and
Joshua N.\ Winn
}

\affil{Department of Physics, and Kavli Institute for
  Astrophysics and Space Research, Massachusetts Institute of
  Technology, Cambridge, MA 02139, USA}

 \journalinfo{Accepted version: 2011 August 26}
 \slugcomment{{\it Astrophysical Journal}, in press}

\begin{abstract}

  We present the analysis of 4 months of {\it Kepler} photometry of the K4V star HAT-P-11, including 26 transits of its ``super-Neptune'' planet. The transit data exhibit numerous anomalies that we interpret as passages of the planet over dark starspots. These spot-crossing anomalies preferentially occur at two specific phases of the transit. These phases can be understood as the intersection points between the transit chord and the active latitudes of the host star, where starspots are most abundant. Based on the measured characteristics of spot-crossing anomalies, and previous observations of the Rossiter-McLaughlin effect, we find two solutions for the stellar obliquity $\psi$ and active latitude $l$: either $\psi = 106^{+15}_{-11}$ and $l=19.7^{+1.5}_{-2.2}$, or $\psi = 97^{+8}_{-4}$ and $l=67^{+2}_{-4}$ (all in degrees). If the active latitude changes with time in analogy with the ``butterfly diagram'' of the Sun's activity cycle, future observations should reveal changes in the preferred phases of spot-crossing anomalies.

\end{abstract}

  
\section{Introduction}

We have been developing a new method to measure the obliquity of a star with respect to the orbital plane of a transiting planet. Obliquities are important because they are fundamental geometric parameters, and because they bear clues about the formation, migration, and tidal evolution of close-in planets (see, e.g., Queloz et al.\ 2000; Ohta et al.\ 2005; Winn et al.\ 2005, 2010a; Fabrycky \& Winn 2009; Triaud et al. 2010; Morton \& Johnson 2011). The traditional method involves observations of the Rossiter-McLaughlin (RM) effect, a spectroscopic anomaly that occurs during transits due to selective blockage of the stellar rotation field (see, e.g., Queloz et al.\ 2000; Ohta et al.\ 2005; Gaudi \& Winn 2007). The new method is purely photometric, and is based on observations of photometric anomalies in transit light curves resulting from the passage of the planet in front of starspots.

When the planet blocks the light coming from a relatively dark portion of the stellar photosphere, the fractional loss of light is temporarily reduced, and a positive ``bump'' is observed in the light curve (see, e.g., Rabus et al.~2009). Sanchis-Ojeda et al.~(2011) showed how the recurrence (or not) of these starspot-crossing anomalies can be used to measure or place bounds upon the stellar obliquity. In simplest terms, the recurrence of an anomaly in two closely-spaced transits is evidence for a low obliquity, because otherwise the spot would rotate away from the transit chord. We and our colleagues applied this technique to the particular system WASP-4, showing that the new method gives stronger constraints on the stellar obliquity than the previous observations of the RM effect (Triaud et al.~2010). 

Independently, Nutzman et al.~(2011) used the recurrences of spot-crossing anomalies (as well as the phase of the out-of-transit modulation of the total light) to confirm a low obliquity for CoRoT-2b. More recently, D{\'e}sert et al.~(2011) found a similar pattern of recurrences for Kepler-17b, and concluded that the host star has a low obliquity.  The pattern in that case was made even more dramatic by the near-integral ratio between the rotational and orbital periods.

Sanchis-Ojeda et al.~(2011) noted that another interesting target for this method would be HAT-P-11, a K4V star with a ``super-Neptune'' planet of mass 26~$M_\oplus$ and radius 4.7~$R_\oplus$ in a 4.9-day orbit. The star was already suspected of having starspots (Bakos et al.~2010), it had been shown to have a high obliquity based on RM observations (Winn et al.~2010b, Hirano et al.~2011), and most importantly, it lies within the field of the view of the {\it Kepler} photometric satellite (Borucki et al.~2010). Several months of nearly continuous, highly precise {\it Kepler} photometry are already available, and several years of data will eventually become available.

In this paper we analyze public data from {\it Kepler}, which have indeed revealed numerous spot-crossing anomalies, and led to constraints on the stellar obliquity, although not in the manner we anticipated.  Section 2 presents the data, and Section 3 gives our estimates for the basic system parameters as well as the times of spot-crossing anomalies. The anomalies occurred preferentially at two particular phases of the transit. Section 4 presents a simple geometric model, in which special phases are the intersection points between the transit chord and the active (starspot-rich) latitudes of the star. Section 5 discusses the results.

While preparing this paper we learned that two other analyses of the public {\it Kepler} data have been undertaken, by Southworth~(2011) and by Deming et al.~(2011). We refer the reader to those works for a different perspectives on the data, focusing on refinement of the transit parameters rather than the spin orientation of the star.

\section{Analysis of out-of-transit data}

We used the Multimission Archive at STScI to obtain the {\it Kepler} data for HAT-P-11 spanning the 140-day interval from 2009~May~02 to September~16, with a time sampling of one minute. In {\it Kepler} parlance, we used {\sc ap\_raw\_flux} short-cadence data from quarters 0, 1, and 2.

\begin{figure*}[!htp]
\begin{center}
\leavevmode
\hbox{
\epsfxsize=6.5in
\epsffile{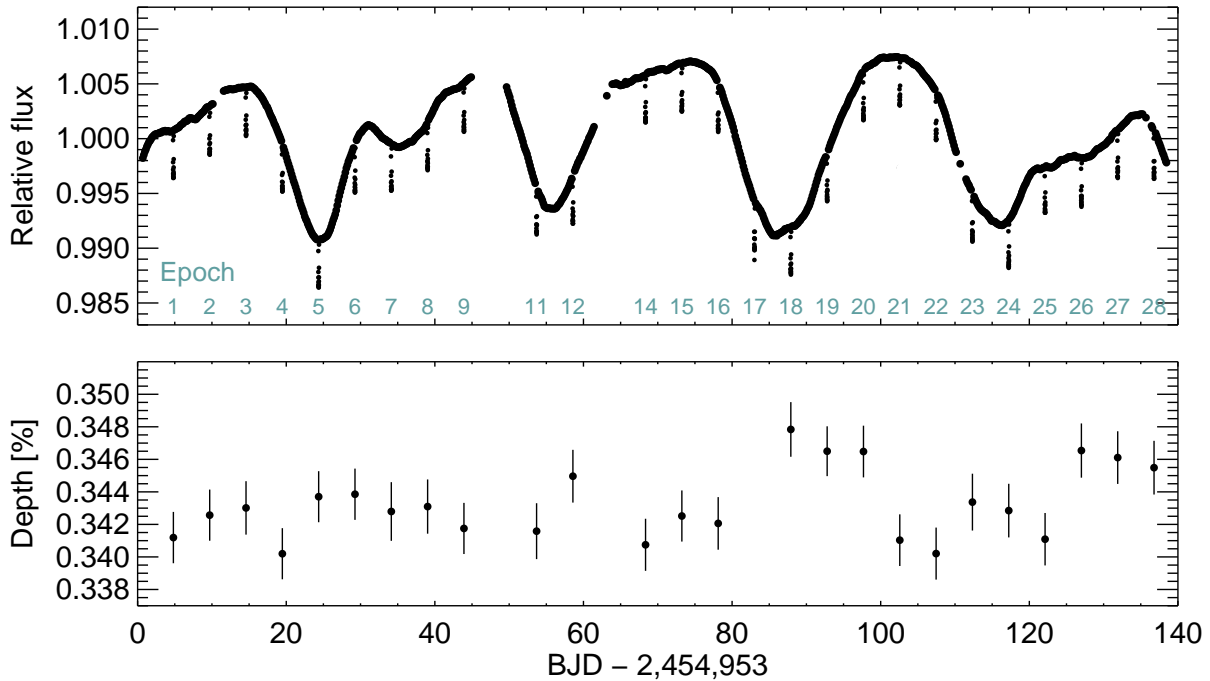}}
\end{center}
\vspace{-0.1in}
\caption{ {\bf {\it Kepler} photometry of HAT-P-11.} {\it Top.}---The time series considered in this paper. The transits are evident as 0.4\% dips with a period of 4.9~days, and are labeled with epoch numbers to facilitate cross-referencing with Figures~3 and 4 and the accompanying text.  Outside of transits there are 4 prominent minima, probably representing a relatively dark starspot pattern being carried around by stellar rotation.  {\it Bottom.}---The measured depth of each transit, using the procedure described in Section 3. The transit depth is defined as the square of the fitted planet-to-star radius ratio.}
\label{fig:totallc}
\vspace{0.1in}
\end{figure*}

The time series is shown in Figure~\ref{fig:totallc}, after expunging a few glaring outliers and normalizing each quarter of data to have a mean of unity.  A total of 26 transits were observed. Two transits were missed due to interruptions in satellite observing. One transit (epoch 17 as identified in Fig.~\ref{fig:totallc}) exhibited irregular flux jumps that were also observed in light curves of nearby stars, thereby implicating an instrumental problem. Data from this transit were excluded from our analysis.

\begin{figure}[!h]
\epsscale{1.0}
\plotone{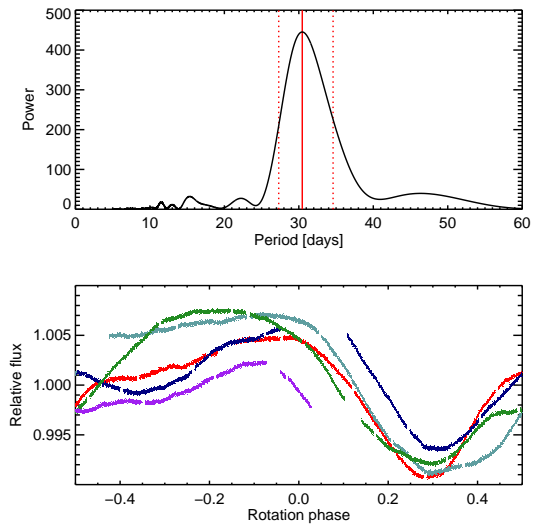}
\caption{ {\bf Rotation period of HAT-P-11.} {\it Top.}---Lomb-Scargle periodogram
of the out-of-transit data. The peak (solid vertical line) is at 30.5~days and
the full-width at half-maximum (dotted vertical lines) span the range
from 27.3 to 34.6~days. {\it Bottom.}---Relative flux as a function
of rotational phase, after folding with a period of 30.5~days. The minima nearly coincide at
a phase of 0.3.}
\label{fig:rotper}
\end{figure}

The relative flux varied by about 1.5\% (peak to peak), with four prominent minima spaced apart by about 30 days. The spacing of the minima probably represents the stellar rotation period. The light curve is not strictly periodic, nor is it expected to be strictly periodic. Among the possible sources of aperiodicity are differential rotation, which causes spots at different latitudes to have differing rotating periods, as well as slow changes in the sizes, shapes, and locations of starspots. The top panel of Figure~\ref{fig:rotper} shows a Lomb-Scargle periodogram of the out-of-transit data. We estimate $P_{\rm rot}=30.5_{-3.2}^{+4.1}$~days, based on the peak and full-width at half-maximum of the periodogram. The bottom panel of Figure~\ref{fig:rotper} shows the out-of-transit data folded with a period of 30.5~days. This is consistent with the earlier estimate of 29.2~days by Bakos et al.~(2010) from ground-based photometry.

\section{Analysis of transit data}

\begin{figure*}[!htp]
\begin{center}
\leavevmode
\hbox{
\epsfxsize=6.5in
\epsffile{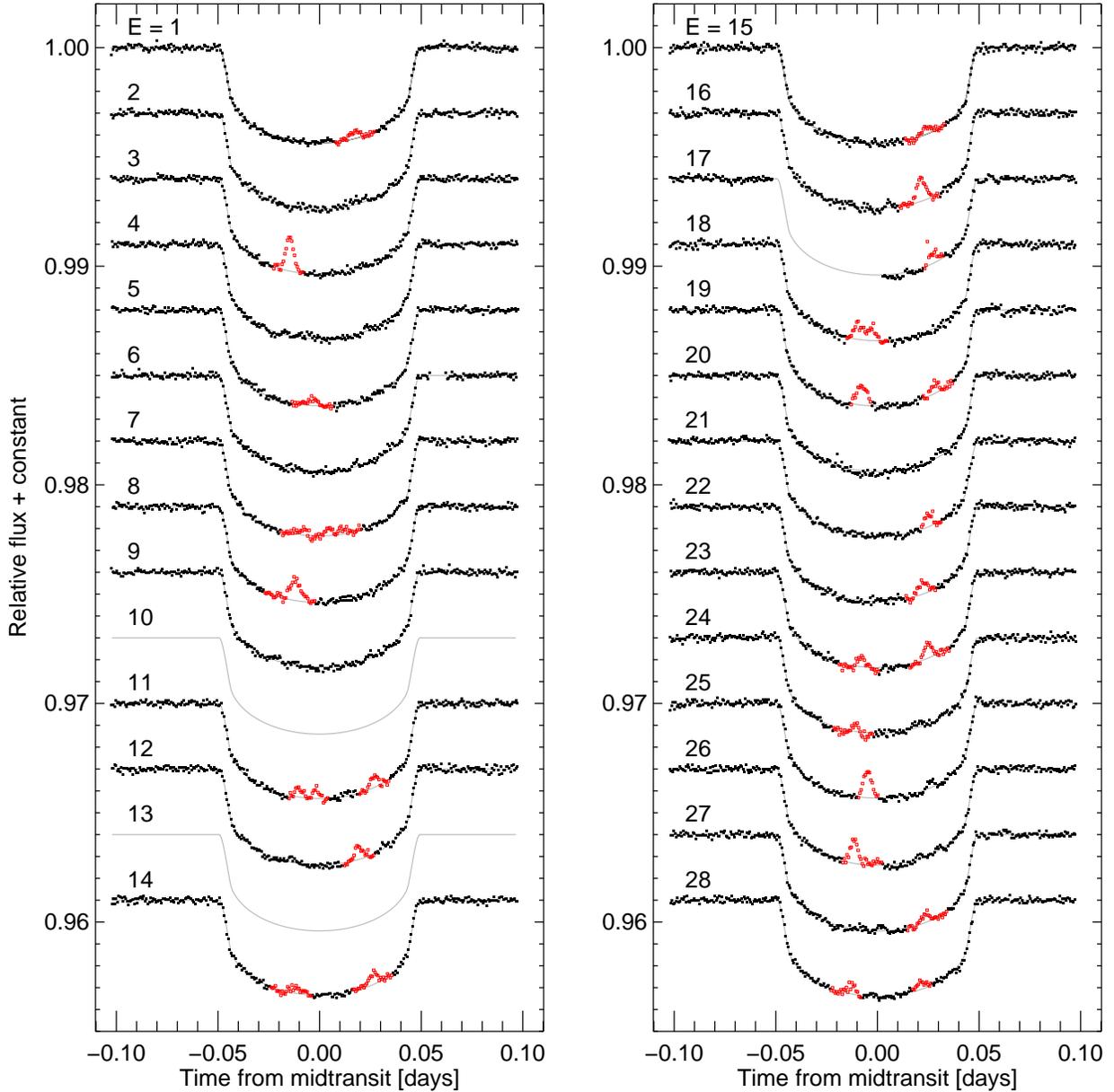}}
\end{center}
\vspace{-0.0in}
\caption{ {\bf {\it Kepler} observations of transits of HAT-P-11.}
Based on data from quarters 0, 1, and 2.
The best-fitting model curves are shown as thin gray lines.
Red squares are points that are suspected of being
strongly affected by spot-crossing events,
and were assigned zero weight in the fitting procedure.}
\label{fig:alltrans}
\vspace{0.1in}
\end{figure*}

To analyze the transits, we selected all the data within 4.8~hr (twice the transit duration) of any predicted transit midpoint. The data surrounding each transit were fitted with a Mandel \& Agol (2002) model, using a quadratic limb-darkening law and allowing for a linear trend in the out-of-transit flux.  Nearly every transit showed an anomaly that was not well-fitted by the transit model, which we interpret as the consequence of irregularities on the stellar photosphere. These anomalies were identified ``by hand'' and assigned zero weight in a subsequent fit to the data. 
Figure~\ref{fig:alltrans} shows the resulting light curves, along with the best-fitting model and the data points that were excluded from the fit.

In this fit, each transit was allowed to have a unique value of the transit depth. The intention was to allow for the possibility of apparent depth variations due to changes in the flux of the untransited portion of the star. The results for the transit depth were plotted in Figure~\ref{fig:totallc}. They are essentially consistent with a constant depth, which in retrospect is not surprising since the stellar flux changes by $\pm$0.8\% and the individual transit depths are measured with a precision of only $\pm$1\%. Table~1 gives the results for the system parameters, based on a Monte Carlo Markov Chain algorithm.  The quoted value of the planet-to-star radius ratio is based on the mean, and the standard error in the mean, of all 26 measured transit depths.

\begin{deluxetable}{lcccc}

\tabletypesize{\scriptsize}
\tablecaption{System parameters of HAT-P-11\label{tbl:parameters}}
\tablewidth{0pt}

\tablehead{
\colhead{Parameter} &
\colhead{Value} &
\colhead{Uncertainty}
}

\startdata
{\it Transit ephemeris:} & & \\
Reference epoch~[BJD]      & $2454957.812464$  &  $0.000022$      \\
Orbital period~[days]            & $4.8878049$      &  $ 0.0000013$    \\
& &  \\
{\it Transit parameters:} & &  \\
Planet-to-star radius ratio, $R_p/R_\star$             & $0.05862$         &  $0.00026$  \\
Transit duration~[days]             & $0.09795$          &  $0.00006$                            \\
Transit ingress or egress duration~[days]         & $0.00550$          &  $0.00007$    \\
Linear limb-darkening coefficient, $u_1$                & $0.599$           &  $0.015$   \\
Quadratic limb-darkening coefficient, $u_2$             & $0.073$           &  $0.016$   \\
Transit impact parameter, $b$        & $0.132$           &  $0.045$   \\
Scaled semimajor axis, $a/R_\star$                     & $15.6$           &  $1.5$
\enddata

\tablecomments{Based on a Markov Chain Monte Carlo analysis of the 26 {\it Kepler} light curves,
with uniform priors on $R_p/R_\star$, $b$, and $a/R_\star$, and Gaussian priors
on the eccentricity parameters $e\cos\omega = 0.201\pm 0.049$ and $e\sin\omega = 0.051\pm 0.092$.
The quoted values and uncertainties are based on the $15.65\%$, $50\%$ and $84.35\%$ levels of the cumulative
distributions of the marginalized posteriors.}
\end{deluxetable}

\section{Analysis of spot-crossing anomalies}

\subsection{Simple test for spin-orbit alignment}

Because so many spot-crossing anomalies were detected, an immediate test is available for spin-orbit alignment. The logic is as follows. If the stellar obliquity were zero ($\psi=0$), then the transit chord would correspond to a certain fixed range of latitudes in the reference frame of the star. In that case, after a given spot-crossing anomaly, that same spot would advance along the transit chord due to stellar rotation and future spot-crossing events can be predicted and sought out in the data. For HAT-P-11, a spot-crossing anomaly observed in the first half of the transit would recur at a later phase of the next transit. This is because the orbital period (4.9~days) is shorter than half a rotation, the time it takes for the spot to cross the visible stellar hemisphere. The underlying assumption is that the spot does not move significantly or fade into undetectability within 4.9~days, but that assumption seems justified (for large spots at least) given the observed coherence of the light curve over 4 rotations (see Figure~\ref{fig:rotper}).

\begin{figure*}[!htp]
\begin{center}
\leavevmode
\hbox{
\epsfxsize=5.0in
\epsffile{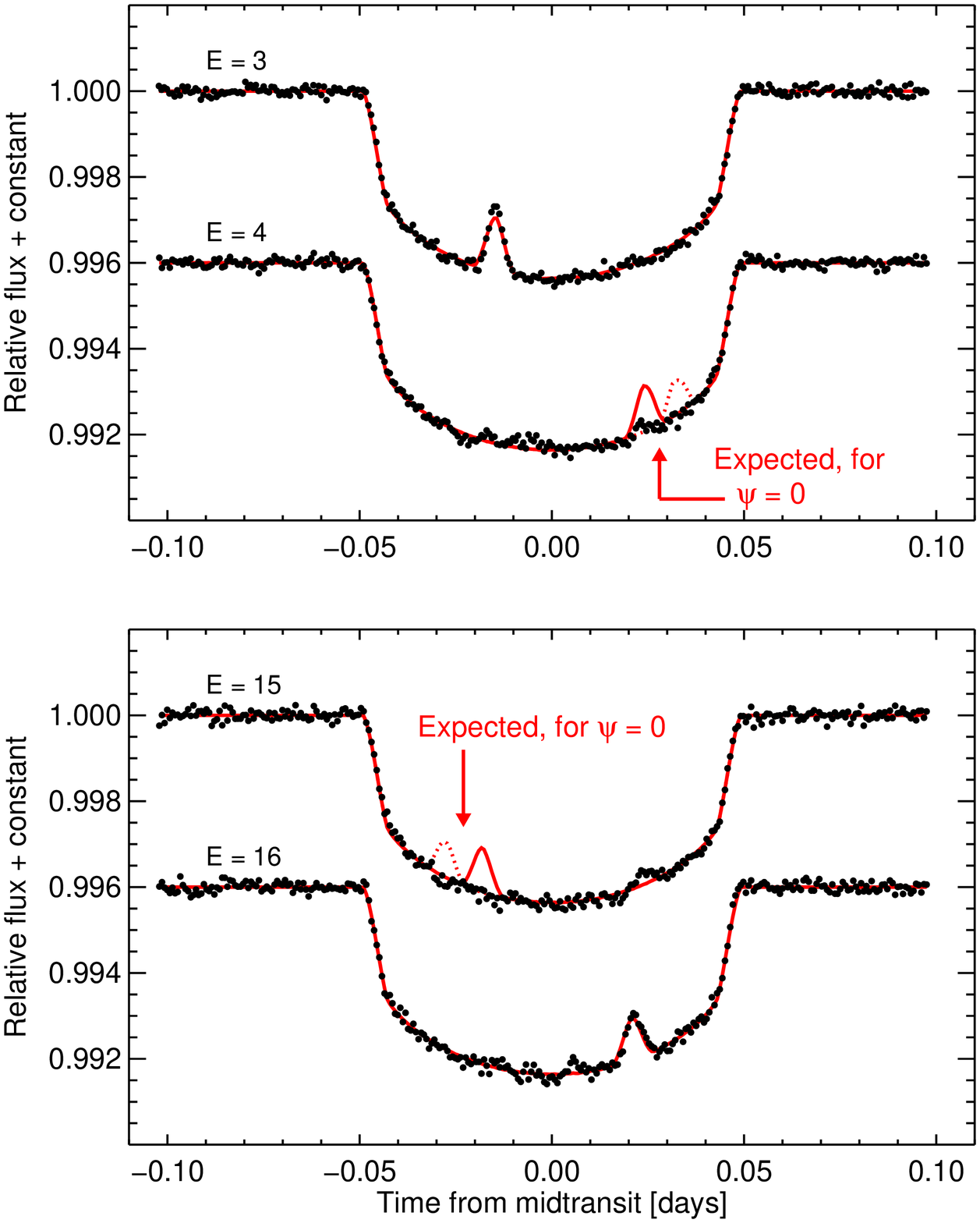}}
\end{center}
\vspace{-0.0in}
\caption{ {\bf Evidence for spin-orbit misalignment.} Shown are two examples of pairs of consecutive transits
where one spot-crossing anomaly was observed, and if $\psi$ were zero, there would
have been a corresponding spot-crossing anomaly detected in the other
transit. No such correspondences were observed in the time series considered in this paper.
The black dots are data points, and the red lines
are best-fitting models including a circular spot with a lower intensity than the surrounding
photosphere. For epochs 4 and 15, two curves for the expected spot signal are plotted (solid and dotted),
corresponding to extremes in the range of rotation periods from 27.3 to 34.6~days.}
\label{fig:spotpred}
\end{figure*}

No such recurrence is seen in the {\it Kepler} data, leading to the conclusion the star's spin axis is misaligned with the planet's orbital axis. Figure~\ref{fig:spotpred} shows two of the clearest examples of a pair of transits where one spot-crossing anomaly was seen, and the other corresponding anomaly that would be predicted for perfect spin-orbit alignment is missing. Many other examples are evident in Figure~\ref{fig:alltrans}.

\subsection{Evidence for spin-orbit misalignment}

Winn et al.~(2010b) suggested that even for $\psi \neq 0$, the recurrence of spot-crossing anomalies could be observed and used to constrain the stellar obliquity and the stellar rotation period.  However, such recurrences require the spots to last for one or more full rotation periods, as opposed to one-sixth of a rotation period, and they also require the rotation period to be a nearly exact multiple of the orbital period.  This latter condition may or may not be the case for HAT-P-11, and is {\it a priori} unlikely. Given the uncertainty in the rotation period, the ratio of rotation to orbital periods is between 5.6 and 7.1. Indeed we could establish no firm correspondences between multiple pairs of spot events.

\begin{figure}[!hpt]
\epsscale{1.2}
\plotone{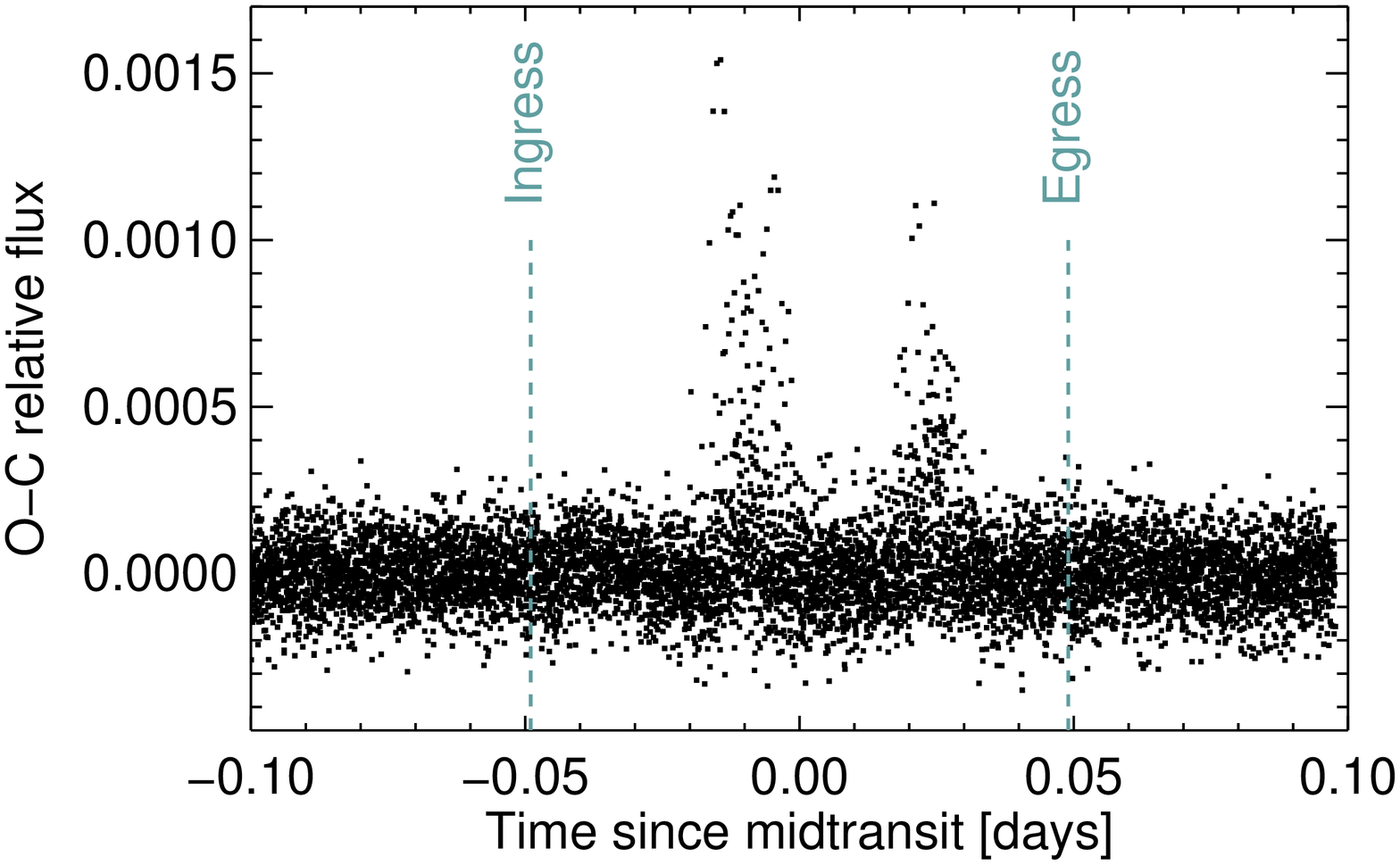}
\caption{ Differences between the data, and the best-fitting transit model,
as a function of transit phase. Data from all 26 transits are plotted.
Spot-crossing anomalies (the large positive residuals)
appear preferentially at two particular phases. These phases are
not symmetrically placed with respect to the transit midpoint.}
\label{fig:allres}
\end{figure}

However, there is a regularity in the pattern of anomalies that we did not anticipate, although perhaps we should have. Figure~\ref{fig:allres} shows the residuals between the data and the best-fitting transit model, as a function of time relative to the nearest midtransit time.  The spot-crossing anomalies are manifested as large positive residuals. They do not occur at random phases of the transit, but rather at two specific phases: approximately $-0.010$ and 0.025~days relative to midtransit.  One might initially interpret this as evidence for two long-lived spots on the star, with each bump representing the intersection of one spot's stellar latitude with the transit chord. However, in that situation one would observe at most two anomalies per rotation period, and more likely fewer, unless the orbital and rotational periods were nearly commensurate. In reality we observed at least 25 anomalies over 4 rotation periods. There are evidently many different spots on HAT-P-11 and they are clustered at two particular stellar latitudes.

If spots appeared with equal probability at any latitude, one would expect to see a nearly uniform distribution of outliers in Figure~\ref{fig:allres}, except near the ingress and egress where limb darkening and geometrical foreshortening would make some spots undetectable. Likewise, if $\psi=0$, then a nearly uniform distribution of residuals would be observed even if the spots were clustered in latitude (again, unless there were some near-commensurability between rotational and orbital periods). Therefore, since the data exhibit two particular peaks, we conclude that the system is misaligned {\it and} that the starspots occur preferentially at certain ``active latitudes.''

The phenomenon of active latitudes is a familiar one from solar astronomy, which is why we wrote above that we should have anticipated this result.  Carrington (1858) and Sp\"{o}rer (1874) found that over the course of the Sun's 11-year activity cycle, the mean latitude of sunspots is sharply defined for any few-month interval, and undergoes a gradual shift from high latitudes to the equator. This spatial regularity of the cycle sometimes called the Sp\"{o}rer law.  The famous ``butterfly diagram'' (Maunder~1904), in which sunspot latitude is charted against time, can be regarded as a graphical depiction of this law.  The regions where sunspots are abundant are well described as relatively narrow bands centered on two particular latitudes placed symmetrically with respect to the solar equator.  Early in a cycle, spots appear at latitudes up to 40 degrees.  As the cycle progresses, new sunspots appear at increasingly lower latitudes, with the last sunspots of a cycle lying close to the equator (Solanki 2003).

\subsection{Geometric model}

For a quantitative analysis of the spot-crossing anomalies, we fitted each anomaly with a simple triangular model with three parameters, the height ($A = $~the amplitude of the anomaly in relative flux units), the width ($\tau=$~total duration) and the midpoint ($t_0=$~the time of the event):
\begin{equation}
F(t) = \left\{ \begin{array}{cc}    A - \frac{2A}{\tau} |t-t_0|  &  |t-t_0| <  \tau/2 \\ 0 & | t-t_0| \geq  \tau/2  \end{array}\right.
\end{equation}
Table~\ref{tbl:spots} gives the best-fitting values of the model parameters for each anomaly. As a measure of statistical significance, the $\Delta\chi^2$ between a no-spot model and the spot model is also given for each event. All the chosen anomalies have $\Delta\chi^2$ exceeding 50.

\begin{deluxetable}{lccccc}

\tabletypesize{\scriptsize}
\tablecaption{Characterization of spot-crossing anomalies\label{tbl:spots}}
\tablewidth{0pt}

\tablehead{
\colhead{Epoch} &
\colhead{Amplitude} &
\colhead{Duration}&
\colhead{BJD$-$}&
\colhead{$\Delta \chi^2$ } &
\colhead{Phase} \\
\colhead{} &
\colhead{[ppm] } &
\colhead{[min]}&
\colhead{$2,454,953$} &
\colhead{} &
\colhead{[$x/R_\star$]}
}

\startdata
1 &386 &13.6 &4.830 &86.7 & 0.37\\
3 &1711 &12.2 &14.573 &1646.7 & $-0.31$ \\
5 &324 &16.8 &24.360 &73.7 & $-$0.08\\
8 &984 &16.9 &39.015&716.0 & $-$0.25\\ 
11 &440 &10.4 &53.680 &121.9 & $-$0.22 \\
11 &530 &7.5 &53.689 &119.4 & $-$0.04\\ 
11 &593 &10.1 &53.717 &191.6 & 0.57\\
12 &660 &12.2 &58.597 &162.0 & 0.41\\
14 &282 &17.3 &68.343 &61.9 & $-$0.24\\
14 &573 &12.7 &68.380 &186.3 & 0.57 \\
15 &361 &13.4 &73.266 &59.9 & 0.52\\
16 &1155 &14.3 &78.151 &551.4 & 0.46\\
17 & 618 &20.7 &83.045 &133.4 & 0.58\\
18 &689 &26.9 &87.898 &310.6 & $-$0.15\\
19 &969 &15.1 &92.785 &578.5 & $-$0.17 \\
19 & 525 &10.8 &92.821 &146.8 & 0.60\\
21 & 627 & 12.4 & 102.594 & 207.5 & 0.54\\
22 & 519 & 13.4 & 107.479 & 179.7 & 0.48\\
23  & 479 & 11.4 & 112.336 & 97.8 & $-$0.17\\
23 & 769 & 11.4 & 112.369 & 248.1 & 0.53\\
25 & 1302 & 13.9 & 122.115 & 656.8 & $-$0.10 \\
26 & 1127 & 11.4 & 126.996 & 581.4 & $-$0.25\\
27 & 444 & 12.5  & 131.919 & 84.8 & 0.50 \\
28 & 446 & 10.7 & 136.771 & 101.1 & $-$0.27 \\
28 & 326 & 18.8 & 136.804 & 84.3 & 0.45
\enddata

\end{deluxetable}

Next we used a simple geometric model to constrain the spin orientation of the star as well as the locations and widths of the active zones. The premise of the model is that each of the two features seen in Figure~\ref{fig:allres} represents an intersection between two strips on the stellar disk: the transit chord, which has a width equal to the planet-to-star radius ratio; and an active zone, a range of latitudes surrounding an active latitude.  The intersection points of the two strips determine the centroids of the two features seen in Figure~\ref{fig:allres}.  The width of the intersection region (which depends on the width of each strip, as well as the angle of intersection) determines the widths of the two features seen in Figure~\ref{fig:allres}.

\begin{figure}[!hpt]
\epsscale{1.2}
\plotone{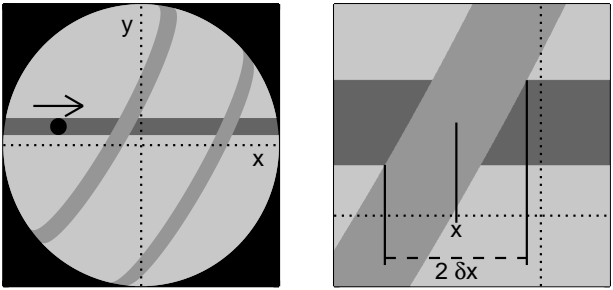}
\caption{{\bf Illustration of the coordinate system.}
The transit chord is parallel to the $x$ axis.
The region of intersection between the transit chord and
an active zone is described by $x$, its center, and
$\delta x$, its width in the $x$-direction.
All distances are expressed in units of the stellar radius.}
\label{fig:sr}
\end{figure}

The first step in the model is to characterize the two features by their central phases and widths. We use a coordinate system centered on the stellar disk, with the $x$-axis aligned with the planet's motion during transits, the $y$-axis in the perpendicular direction in the sky plane, and the $z$-axis pointing along the line of sight (see Figure~\ref{fig:sr}). All distances are measured in units of the stellar radius. The goal of this step is to determine the central values of $x$ for the two spot-anomaly features, which we denote $\bar{x}_1$ and $\bar{x}_2$, and the Gaussian widths of the features, which we denote $\bar{\sigma}_1$ and $\bar{\sigma}_2$.

\vspace{0.2in}
To estimate $\bar{x}$, and $\bar{\sigma}$ for each of the two
features, we maximized the likelihood
\begin{equation}
\mathcal{L} \propto \prod_{i=1}^N~
\int dx~
\frac{1}{\sqrt{2\pi\sigma_i^2}}~
e^{-\frac{(x-x_i)^2}{2\sigma_i^2}}~
\frac{1}{\sqrt{2\pi\bar{\sigma}^2}}~
e^{-\frac{(x_i-\bar{x})^2}{2\bar{\sigma}^2}},
\end{equation}
where the index $i$ runs over all spots contributing to the feature, the first exponential within the integral represents the probability distribution for each measurement of a spot location, and the second exponential is the assumed Gaussian model of the spatial distribution of spots surrounding the active latitude. Loosely speaking, this is similar to fitting a Gaussian function to each feature seen in Figure~\ref{fig:allres}, after transforming the time
coordinate into the $x$ coordinate. The results, based on the same MCMC algorithm used to estimate the system parameters, are $\bar{x}_1 = -0.19 \pm 0.03$, $\bar{x}_2 = 0.51 \pm 0.02$, $\bar{\sigma}_1 = 0.09 \pm 0.02$ and $\bar{\sigma}_2 = 0.07 \pm 0.02$.The quoted values and uncertainties are based on the 15.65\%, 50\% and 84.35\% levels of the cumulative distributions of the marginalized posteriors, distributions that all follow a Gaussian profile. 

\begin{figure*}[!htp]
\begin{center}
\leavevmode
\hbox{
\epsfxsize=6.5in
\epsffile{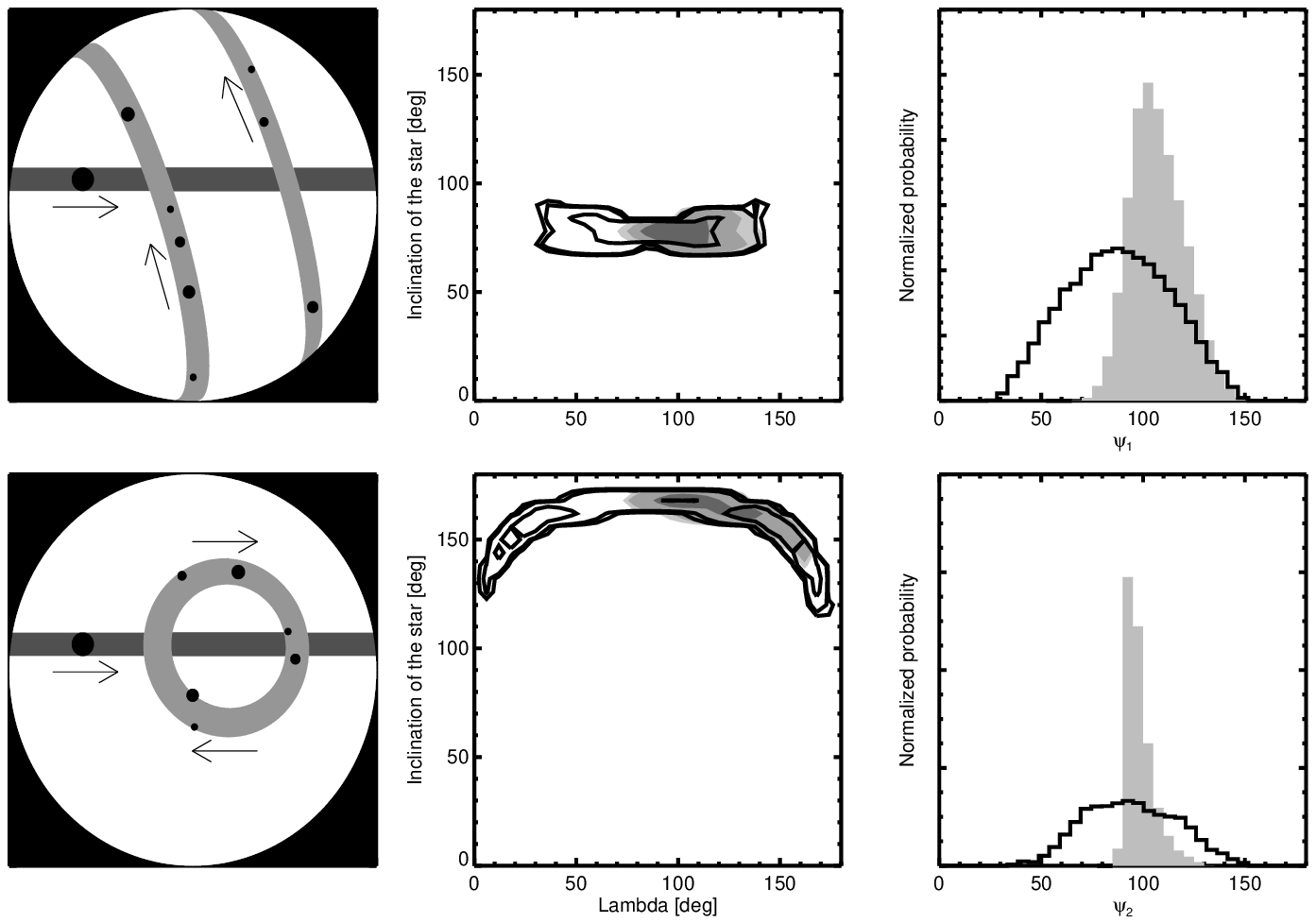}}
\end{center}
\vspace{-0.1in}
\caption{{\bf Two solutions for the stellar geometry},
and associated results of parameter estimation.
The upper panels represent the double-banded, edge-on solution
and the lower panels represent the single-band, pole-on solution.
The left column of panels are
sketches of the system, using the most probable values of the parameters.
The central column shows two-dimensional
posterior distributions for the stellar orientation parameters $\lambda$ and $i_s$,
with solid lines representing the results with a uniform prior on $\lambda$,
and gray scales for the results with a two-sided Gaussian prior
$\lambda = 103^{+26}_{-10}$~deg based on the RM results of Winn et al.~(2010b).
The confidence levels are 68.3\%, 95\%, and 99.73\%.
The right column shows the posterior
distribution for the true spin-orbit angle $\psi$, again with
the solid line representing the result from the spot analysis alone,
and the shaded distribution representing the joint
results of the spot analysis and the RM measurement.}
\label{fig:mcmc}
\vspace{0.1in}
\end{figure*}

The next step is to use these results to constrain the spin orientation of the star and the locations of the active zones. The spin orientation of the star is parameterized by the inclination angle $i_s$ of the north pole with respect to the line of sight, and the sky-plane angle $\lambda$ between the north pole and the orbit normal (i.e., the same angle probed by observations of the RM effect).  In analogy with the Sun, and based on the overall symmetry of the star, we assume the active zones to be symmetrically placed with respect to the stellar equator, centered on latitudes $\pm l$ and with half-widths $\delta l$. Therefore our next task is to use spherical trigonometry to relate $\bar{x}_1$, $\bar{x}_2$, $\bar{\sigma}_1$ and $\bar{\sigma}_2$ to $i_s$, $\lambda$, $l$ and $\delta l$.

The geometry is illustrated in Figure~\ref{fig:sr}.
A point $\vec{r} \equiv (x,y,z)$ on the middle of the transit chord has $y=b$ (where $b$ is the impact parameter), $z>0$ and $x^2 + y^2 + z^2 =1$.
The stellar north pole is at (following the convention of Ohta et al.~2005)
\begin{equation}
\vec{r}_{{\rm NP}} = -\sin{i_s}\sin{\lambda}~\hat{x} + 
\sin{i_s}\cos{\lambda}~\hat{y} + 
\cos{i_s}~\hat{z}.
\end{equation}
The latitude $l$ of a given point on the surface of the star can be calculated given its $xyz$ coordinates and the position of the north pole:
\begin{equation}
l = \frac{\pi}{2} - \cos^{-1}{\left(1-d^2/2\right)}, \qquad d^2 \equiv (\vec{r}-\vec{r}_{{\rm NP}})^2.
\end{equation}
These relations allow us (after some algebra) to
compute the $x$ coordinates of any intersection points
of the transit chord and active latitude:
\begin{equation}
x_{1,2}(\lambda, i_s, b, l) =
\frac{(b\sin{i_s}\cos{\lambda}-\sin{l})\sin{i_s}\sin{\lambda}\pm \sqrt{f(\lambda, i_s, b, l)}}{1-\sin^2{i_s}\cos^2{\lambda}}.
\label{eq:xcoord}
\end{equation}
Here, the factor $f(\lambda, i_s, b, l)$, defined as
\begin{equation}
\cos^2{i_s}[1-\sin^2{i_s}\cos^2{\lambda}-b^2+2b\sin{l}\sin{i_s}\cos{\lambda}-\sin^2{l}],
\end{equation}
determines whether there are zero, one or two different intersection points for a given set of values of $(\lambda, i_s, b, l)$.  Since we are assuming there are two active latitudes $\pm l$, we can have up to four intersection points.  Once all possible solutions have been found, it is necessary to check that they occur on the visible hemisphere of the star ($z>0$).

\begin{figure*}[!htp]
\begin{center}
\leavevmode
\hbox{
\epsfxsize=6.5in
\epsffile{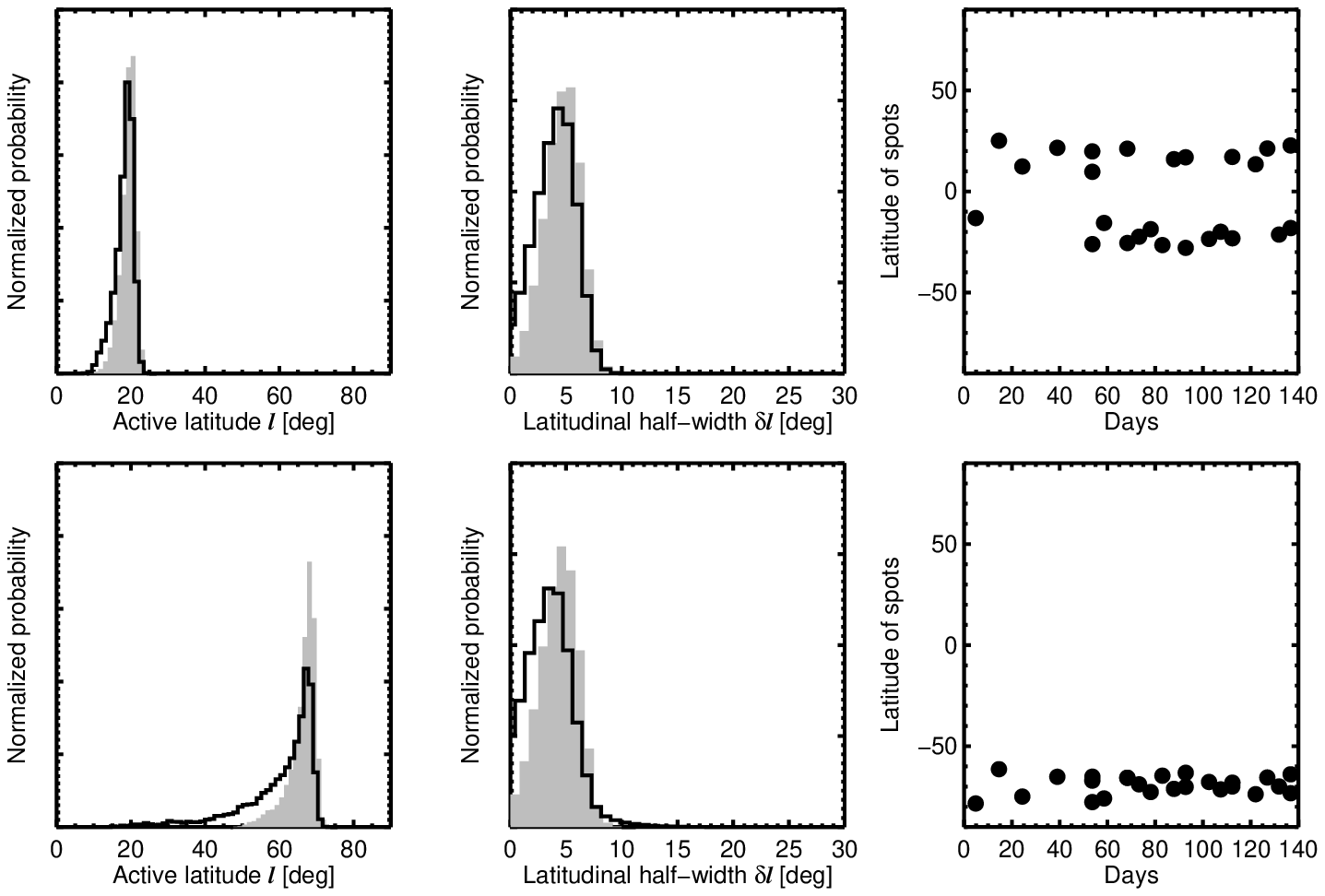}}
\end{center}
\vspace{-0.1in}
\caption{{\bf Results for the parameters describing the active zones.}
Upper panels represent the two-band, edge-on
solution and lower panels represent the single-band, pole-on solution.
On the left are posteriors for $l$, the central latitude of the active region.
(In the model, the active latitudes are symmetrically placed at $\pm l$.)
As in Figure~\ref{fig:mcmc}, the solid line is for the case of a uniform
prior in $\lambda$, and the shaded histogram is
for the case when the RM result for $\lambda$ was used as a prior.
The central column shows the corresponding results for the parameter $\delta l$
describing the latitudinal half-width of the active regions.
The right column shows the butterfly diagram for HAT-P-11, based on the measured phases
of the spot anomalies and the best-fitting
values of $i_s$ and $\lambda$. }
\label{fig:mcmcextra}
\vspace{0.1in}
\end{figure*}

The impact parameter $b$ has already been determined from transit photometry.  We can adjust the other parameters $\lambda$, $i_s$ and $l$ to achieve a match between the $x$-values of the intersection points and the previously determined values of $\bar{x}_1$ and $\bar{x}_2$.  There are four solutions, two of which are illustrated in the left panels of Figure~\ref{fig:mcmc}.  One solution has the active zones at relatively low latitudes, with both the northern and southern active latitudes intersecting the transit chord.  The second solution has the star oriented nearly pole-on, so that one active latitude presents an ellipse on the sky that intersects the transit chord twice.  The third and fourth solutions are related to the first two by the symmetry $\{ \lambda, i_s, l\} \rightarrow \{ \pi+\lambda, \pi-i_s, l\} $. These ``mirror solutions'' give similar results for the three-dimensional obliquity $\psi$ as the other two solutions. However since they give negative values of $\lambda$ that are ruled out by observations of the RM effect (Winn et al.~2010b, Hirano et al.~2011) for brevity we do not discuss them in the rest of this paper.

Next we make use of the measured widths of the spot-anomaly distributions.  Using Eqn.~(\ref{eq:xcoord}), we may calculate the four vertices of the intersection region between the transit chord and the band of active latitudes.  We then take the difference between the maximum and the minimum values of $x$ and divide by two, a quantity we will call $\delta x$.  By characterizing the width in this manner, we are effectively assuming that spots are equally likely to form anywhere in the range $l \pm \delta l$. This is computationally very convenient, but it is in mild contradiction with the Gaussian distribution we adopted when estimating $\bar{\sigma}_1$ and $\bar{\sigma}_2$. For this reason, we multiply $\delta x$ by $\sqrt{2/\pi}$, to give the standard deviation of a Gaussian function that has the same area as a uniform distribution with width $\delta x$.  This can then be compared with $\bar{\sigma}_1$ and $\bar{\sigma}_2$, using a goodness-of-fit function,
\begin{eqnarray}
\chi^2 ( \lambda, i_s, l, \delta l, b) & = & 
\left(\frac{\bar{x}_{1,~{\rm calc}}-(-0.19)}{0.03}\right)^2 + \left(\frac{\bar{x}_{2,~{\rm calc}}-(0.51)}{0.02}\right)^2 \nonumber \\
& + & \left(\frac{\sqrt{2/\pi}~\delta \bar{x}_{1,~{\rm calc}}-0.09}{0.02}\right)^2 \nonumber \\
& + & \left(\frac{\sqrt{2/\pi}~\delta \bar{x}_{2,~{\rm calc}}-0.07}{0.02}\right)^2,
\end{eqnarray}
where ``calc'' indicates the value that is calculated geometrically based on the parameters $\lambda$, $i_s$, $l$, $\delta l$, and $b$.  We thereby enforce agreement between the ``observed'' and ``calculated'' properties of the intersection region (their locations and widths), under the assumption that the uncertainties in the 'observed' quantities are Gaussian-distributed.

To explore the parameter space we used an MCMC technique, with the Gibbs sampler and the Metropolis-Hastings algorithm, and a likelihood proportional to $\exp(-\chi^2/2)$.  Uniform priors were adopted on all parameters except the impact parameter, for which a Gaussian prior was adopted based on the light curve analysis presented earlier ($b=0.132\pm 0.045$), and the stellar inclination angle $i_s$, for which the prior was uniform in $\cos i_s$ (i.e., isotropic). In addition, we considered two cases: one in which the prior was uniform in $\lambda$, and another in which a two-sided Gaussian prior was adopted to enforce agreement with the RM result $\lambda = 103^{+26}_{-10}$~deg (Winn et al.~2010b).

Figure~\ref{fig:mcmc} shows the results, both as contour plots in the space of the stellar orientation parameters $i_s$ and $\lambda$, and as marginalized distributions for the true obliquity $\psi$ that is computed from those two parameters:
\begin{equation}
\cos{\psi} = \cos{i_s}\cos{i_o}+\sin{i_s}\cos{\lambda}\sin{i_o},
\end{equation}
where $i_o$ is the orbital inclination. As discussed earlier, there are two families of solutions. In one solution, the star's equator is viewed nearly edge-on, while in the second solution, the star is viewed nearly pole-on. Both solutions are compatible with the RM results. When the previous RM results are used as a prior, the results are sharpened. The spot-crossing analysis and the RM analysis are complementary in the sense that the RM analysis is sensitive only to $\lambda$, while the spot-crossing analysis is sensitive to a complicated combination of $i_s$ and $\lambda$.  Although the two families of solutions are quite different in the alignment of the star with respect to Earth, they are similar in that they both represent strongly misaligned systems with $\psi \approx 90^\circ$.

Figure~\ref{fig:mcmcextra} shows the results for $l$ and $\delta l$, the parameters describing the active zones on the star. The edge-on solution has a lower value of $l$ than the pole-on solution, but they both give similar results for the latitudinal half-width $\delta l$ of the active regions. The fitted half-widths are around $5^\circ$, which is similar to the observed widths of the active bands on the Sun. The latitude $l\approx 20^\circ$ of the edge-on solution is also a good match to the Sun, while the higher latitude of $l\approx 60^\circ$ for the pole-on solution would be atypically high for the Sun.

For each solution we also plot the analog of the solar butterfly diagram: the latitude of each individual observed spot anomaly as a function of time. We do this in the right column of Figure~\ref{fig:mcmcextra}, based on the best-fitting model in each of the two families.  In the first case, spots cover two symmetric ranges of latitudes, whereas in the second case (the pole-on solution) we see only one band. Eventually, after several more years of {\it Kepler} data are collected, we might expect this type of diagram to show time variations in the latitudes of spots due to the stellar activity cycle on HAT-P-11. This would be a valuable opportunity to construct a butterfly diagram for a star other than the Sun (see also Berdyugina \& Henry 2007).

\section{Discussion}

\begin{deluxetable*}{lcccc}
\tabletypesize{\scriptsize}
\tablecaption{Results of MCMC analysis\label{tbl:mcmc}}
\tablewidth{0pt}

\tablehead{
\colhead{Parameter} &
\colhead{Edge-on solution} &
\colhead{Edge-on solution} &
\colhead{Pole-on solution} &
\colhead{Pole-on solution} \\
\colhead{} &
\colhead{} &
\colhead{($+$~RM prior)} &
\colhead{} &
\colhead{($+$~RM prior)}
}

\startdata
Projected obliquity, $\lambda$~[deg]  & $90\pm 28$ & $106^{+15}_{-12}$  & $83^{+77}_{-65}$&   $121^{+24}_{-21}$ \\
Stellar inclination, $i_s$~[deg]  & $80^{+5}_{-3}$  & $80^{+4}_{-3}$   & $160^{+9}_{-19}$  & $168^{+2}_{-5}$ \\
Obliquity, $\psi$~[deg]  & $91  \pm 27$ &  $106^{+15}_{-11}$ &  $89^{+27}_{-27}$ & $97^{+8}_{-4}$ \\
Latitude of active zone~[deg]  & $19.3^{+1.7}_{-3.0}$  & $19.7^{+1.5}_{-2.2}$ & $63^{+6}_{-17}$  & $67^{+2}_{-4}$  \\
Half-width of active zone~[deg] & $4.5^{+1.7}_{-2.1}$ &$4.8^{+1.5}_{-1.8}$  & $4.0^{+2.3}_{-2.2}$  & $4.5^{+1.6}_{-1.9}$
\enddata
\end{deluxetable*}

The main results of our study are (1) the finding that the starspots on HAT-P-11 are preferentially found at certain active latitudes; (2) the confirmation that the HAT-P-11 star is misaligned with the orbit of its close-in planet; (3) the placement of quantitative bounds on the three-dimensional stellar obliquity based on the observed pattern of spot anomalies and a simple geometrical model. Two families of geometric solutions were found, one in which the star is viewed nearly pole-on, and the other in which the star is viewed closer to edge-on.  Even though both solutions agree on the stellar obliquity, they differ in the locations of the active latitudes: one solution gives more Sun-like active latitudes of about $20^\circ$ while the other solution favors a larger latitude of about $60^\circ$. Breaking this degeneracy would therefore tell us whether or not the activity cycle on HAT-P-11 resembles that of the Sun in this respect.

One might think that the measurement of $v\sin{i_s}$ would help to distinguish these two cases, but the two solutions predict values of $v\sin{i_s} = (2\pi R_\star/P_{\rm rot}) \sin{i_s}$ equal to $1.3$ km~s$^{-1}$ and $0.5$~km~s$^{-1}$, which are both compatible with the observed value of $1.5 \pm 1.5$ km~s$^{-1}$ (Bakos et al.~2010). A basic difference between the pole-on and edge-on solutions is that the pole-on solution would produce smaller out-of-transit (OOT) flux variations for a spot of a given size and intensity.  Indeed, an exactly pole-on solution would not produce any OOT variations. For nearly pole-on configurations, the OOT variations arise only from small variations in limb darkening and geometrical foreshortening along the nearly-circular trajectory of the spot on the stellar disk.  In contrast, for the edge-on solution, the spot disappears from view for half of the rotation period, resulting in larger variations.

Our geometric model did not make use of the information borne by the out-of-transit (OOT) flux variations; could this be used to break the modeling degeneracy and sharpen the constraints?  We have made some efforts in this direction, but they have been inconclusive, mainly because the spot sizes and intensties are not known {\it a priori}. Our simulations show that it is possible that the pole-on solution is correct, and that the $1.5\%$ OOT variations are produced by large and/or dark spots.  Another feature of the pole-on model is that each spot crosses the transit chord twice per rotation period, which might lead to a detectable correlation between the observed spot anomalies. We explored this possibility, but could not reach a firm conclusion because a given spot may traverse the transit chord quickly enough that the planet does not always cross it. Nevertheless, future observations with {\it Kepler} and a more detailed spot-by-spot model might one day be used to break the degeneracy.

Future observations may also reveal the changes in the active latitudes that are analogous to the equatorial drift of the active latitudes on the Sun. As is well known, over the 11-year solar activity cycle the active latitudes migrate toward the equator, where they disappear and are then recreated at higher latitudes, the phenomenon that underlies the butterfly diagram.  Should this also occur for HAT-P-11, we would observe a drift in the phases of the spot-crossing anomalies.  Assuming the active latitudes migrate toward the equator, the phases would behave differently for the pole-on and edge-on solutions. In the case of the pole-on solution, the phases will separate apart and move toward the extremes of the transit, while for the edge-on solution, the phases will move closer to one another.  Of course, we should not necessarily assume that the active latitudes migrate to lower latitudes, as is seen on the Sun, but the data themselves may be able to confirm this fact, making use of differential rotation. In fact, Messina \& Guinan (2003) studied latitude migration on young solar analogues through differential rotation and found that for some stars the active latitudes seem to move poleward rather than migrating toward the equator. A combination of the measurement of the drift of the phases and the change in the period will give us a way of breaking the degeneracy and studying the rate of migration of the active latitudes. (For a review of starspots and the various techniques by which they are observed, see Strassmeier~2009.)

Independently of the activity cycle of HAT-P-11, the confirmation that the star has a high obliquity is helpful for understanding the origin of the close-in planet HAT-P-11b. There has been a recent resurgence of the theory that close-in planets are emplaced by few-body gravitational interactions followed by tidal capture, rather than by gradual inspiral due to interactions with a protoplanetary disk (see, e.g., Fabrycky \& Tremaine 2007, Nagasawa et al.~2008, Matsumura et al.~2010). Part of the evidence is the large incidence of high obliquities among stars that host close-in giant planets (see, e.g., Triaud et al.~2010). With our corroboration of the RM results by Winn et al.~(2010b) and Hirano et al.~(2011), HAT-P-11 can be firmly added to this roster.

\acknowledgements We thank Lucianne Walkowicz, Elisabeth Adams, Drake Deming, Andrew Howard, John Johnson, Mercedes Lopez-Morales, Dan Fabrycky, Philip Nutzman, and Josh Carter for stimulating conversations on this topic. We are grateful for financial support from the NASA Origins program (grant no.\ NNX11AG85G). R.S.\ was also supported by a Fellowship Grant for Post-Graduate Studies from the ``la Caixa'' Foundation in Barcelona, Spain.

The data presented in this paper were obtained from the Multimission Archive at the Space Telescope Science Institute (MAST). STScI is operated by the Association of Universities for Research in Astronomy, Inc., under NASA contract NAS5-26555. Support for MAST for non-HST data is provided by the NASA Office of Space Science via grant NAG5-7584 and by other grants and contracts.

\end{document}